\documentclass[12pt]{article}

\usepackage{amsmath,amssymb,amsthm,epsf,epsfig}
\usepackage{amsmath,amssymb}

\begin{document}
\begin{center}
\large{\textbf{RELATIVISTIC SPINNING PARTICLE IN A NON-COMMUTATIVE EXTENDED SPACETIME}}\\
\end{center}
\begin{center}
Sudipta Das {\footnote{E-mail: sudipta.das\_r@isical.ac.in} and
Subir Ghosh {\footnote{E-mail: sghosh@isical.ac.in}\\
Physics and Applied Mathematics Unit, Indian Statistical
Institute\\ 203 B.T.Road, Kolkata 700108, India}}
\end{center}

\vskip 1cm {\it{\textbf{Abstract}}}:\\
The  relativistic spinning particle model, proposed in [3,4], is
analyzed in a Hamiltonian framework. The spin is simulated by
extending the configuration space by introducing a light-like four
vector degree of freedom. The model is heavily constrained and
constraint analysis, in the Dirac scheme, is both novel and
instructive. Our major finding is an associated novel
non-commutative  structure in the extended space. This is obtained
in a particular gauge. The model possesses a large gauge freedom
and hence a judicious choice of gauge becomes imperative. The
gauge fixed system in reduced phase space simplifies considerably
for further study. We have shown that this non-commutative phase
space algebra is essential in revealing the spin effects in the
particle model through the Lorentz generator and Hamiltonian
equations of motion.
\section {\bf {Introduction:}}
A Relativistic Spinning Particle Model (RSPM), even in the
classical context, is quite hard to construct (see for example
\cite{cor}). One of the earlier examples in physical
$3+1$-dimensions, (that we  have extensively used in our work
\cite{sg1} in $2+1$-dimensional anyons), is that of Hanson and
Regge \cite{hr}. The model \cite{hr} had a drawback (indeed
depending on one's point of view) in that the rest mass and spin
parameters are related in a complicated way, {\it{i.e.}} the mass
and spin are {\it{not}} independent. (This problem was resolved
later by one of us \cite{sg1} in $2+1$-dimensions giving rise to
the spinning particle model for anyon. For alternative
formulations of relativistic spinning particle models see
\cite{plus}.) In this perspective it is very significant that
another RSPM was proposed some years ago by Kuzenko, Lyakhovich
and Segal \cite{kuz} and rediscovered independently by Staruszkiewicz
\cite{st}. The scheme of \cite{kuz,st} is based on Wigner's idea
\cite{wig} that quantum mechanical systems ought to be classified
in terms of unitary irreducible representations of Poincare group.
In a classical setup, this can be interpreted \cite{kuz,st} as
having Casimir invariants that are restricted to fixed numerical
values ({\it{i.e.}} they have to be parameters of the model rather
than constants of motion).

The essential idea behind the present approach \cite{kuz,st} is to
generate the particle spin through the introduction of additional
degrees of freedom, in the form of a light-like four vector $k_\mu
(\tau)$. This dynamical variable has the same status as the
particle position $x_\mu (\tau)$. Quite clearly this is an example
of enlarging the  configuration space to a bosonic ``superspace".
In fact fermionic superspace construction to describe spinning
particles was studied much earlier \cite{susy}.

In the present context of spinning particle it is natural to
choose the two Casimirs as
\begin{equation}
P^\mu P_\mu =m^2~;~~W^\mu W_\mu =-(l^2m^2)/4~;~~W_\mu
=-(1/2)\epsilon_{\mu\nu\rho\sigma }M^{\nu\rho}P^\sigma ,
\label{pw}
\end{equation}
with $P_\mu $ and $ W_\mu $ being the canonical momentum and
Pauli-Lubanski pseudovector respectively. $m$ and $l$ are the mass
and spin parameters of the theory but infact $l$ has the dimension
of length. It was established in \cite{kuz,st} and
later in more general context by Kassandrov et.al \cite{wi} that
(\ref{pw}) can lead to an unambiguous form of action (we will come
to it presently) from which $P_\mu$ and $W_\mu $ can be computed
and the identities in (\ref{pw}) checked. Subsequently the authors
of \cite{wi} have carried out a Lagrangian analysis of this
particular RSPM and have come up with interesting observations.

In this perspective our aim is to make an exhaustive Hamiltonian
analysis of the RSPM proposed in \cite{kuz,st} in a gauge fixed framework. The NC spacetime
appears in our reduced phase space. It should be mentioned that in \cite{kuz} a
Hamiltonian analysis was performed in the extended phase space and hence the NC
nature of the spacetime did not show up. The reasons of our study are
primarily twofold but connected. \\
Reason (I): The beauty of this model \cite{kuz,st} (along with the
previous RSPMs \cite{hr,sg1}, for other models having similar
features see {\it{eg}}. \cite{luk}) is that all of them possess a
non-trivial symplectic structure which induces Non-Commutative
(NC) spacetime (or in the more general context NC phase space) and
this noncommutativity induces the richer dynamics that can
incorporate the spin. {\it{Presence of other numerical parameters,
besides the mass, in a free particle Lagrangian is a signal for
this where eventually the other parameter plays the role of NC
parameter.}} Different forms of NC algebra have appeared in the
above mentioned RSPMs \cite{hr,sg1,luk}. Hence from our previous
experience we could make an educated guess that the RSPM of
\cite{kuz,st} should also have an NC phase space \cite{hor},
which, incidentally, will be completely new since the peculiar
symplectic structure of this form has not been explored before.
However, as it turned out, the constraint analysis of this model
\cite{kuz,st}, in the Hamiltonian formulation of Dirac \cite{dir},
is quite non-trivial and interesting. As we will show later, the
model possesses a large amount of gauge freedom and hence a
judicious gauge choice becomes necessary. Our particular system of
gauges is dictated by the natural requirement that we end up with
a (albeit NC) phase space where the Lorentz generators can be
defined unambiguously ensuring that the vectors transform in a
covariant manner. Furthermore, the gauge fixed Hamiltonian reduces
to that of a canonical relativistic spinless particle, (as in the
case of \cite{hr,sg1}), and spin effects in the dynamics are
generated through the NC Dirac bracket algebra. We will elaborate
on this when we get down to the actual computation.\\
Reason (II):  Kassandrov et.al. \cite{wi} made an intriguing
comment that for appropriate choice of the parameters, in
particular identifying $l$ as the Compton length $i=\hbar /(mc)$
the particle spin can be $\hbar/2$. Now, it will be really
worthwhile to attempt a quantization of the model and that will
require a Hamiltonian formulation, symplectic structures, etc.. In
this sense, the present work is a stepping stone for quantizing
this spinning particle.

The paper is organized as follows: In {\bf{Section 2}} we will
introduce the Lagrangian  of RSPM \cite{kuz,st} and reveal the
constraints in the theory. It will be seen that we need to
introduce auxiliary fields in order to reduce the model to a form
that is amenable to Hamiltonian constraint analysis. For
completeness a brief outline of the Dirac constraint analysis
\cite{dir} will be provided. {\bf{Section 3}} will be devoted to
the study of the symplectic structure in a particular set of
gauges so that in the reduced phase space the NC phase space
algebra will be revealed. As we have mentioned, the problem of
gauge fixing is quite non-trivial. In {\bf{Section 4}} the
dynamics of the model will be studied in this particular gauge
where the NC phase space algebra will be used explicitly.
Throughout we will show the correspondence between our results
(obtained in the Hamiltonian framework) to that of the same model
of \cite{wi} (computed in the Lagrangian framework). We will
conclude the paper in {\bf{Section 5}} with future prospects.

\section {\bf{Relativistic Spinning Particle Model:}}
The action for the RSPM, as stated in \cite{st} is,
\begin{equation}
S=\int L d\tau = -m \int d\tau \sqrt{\dot{x}^2} \sqrt{1+l\sqrt{
\frac{(-\dot{k}^2)}{(k \dot x)^2}}} +\int d\tau \lambda k^2.
\label{lag}
\end{equation}
As we pointed out in the Introduction, the action $L$ has two
parameters $m$, the particle mass, and $l$ the spin parameter,
having dimension of length in our system of units. As it happens
in these type of models \cite{hr,sg1,luk} $l$ will play the role
of the NC parameter; $k_\mu $ is a lightlike vector, $k^\mu k_\mu
=k^2 = 0$. We use a shorthand notation $(ab)=a^\mu b_\mu $ and the metric $g_{\mu\nu}=diag(1,-1,-1,-1)$.

The Lagrangian (\ref{lag}) can be considered as a nontrivial
extension of the Nambu-Goto form of spinless particle, to which
(\ref{lag}) reduces to for $l=0$, since the $\lambda k^2$ term is
non-dynamical and gets decoupled from the dynamical term $-m
\sqrt{\dot{x}^2}$ . One can directly define the conjugate momenta
and angular momentum \cite{wi} $$ P_\mu =\frac{\partial
L}{\partial \dot{x}^\mu}~;~Q_\mu =\frac{\partial L}{\partial
\dot{k}^\mu}~;~M_{\mu\nu }=x_\mu P_\nu -x_\nu P_\mu +k_\mu Q_\nu
-k_\nu Q_\mu $$ and check explicitly that the Casimir relations
(\ref{pw}) are satisfied.

However for Hamiltonian analysis, this form of the model, with its
involved time derivative structure and presence of nested square
roots, is simply not suitable because one will immediately run
into trouble in trying to express the velocities $\dot x_\mu $ and
$\dot k_\mu $ in terms of the momenta $P_\mu $ and $Q_\mu $. For
this we need to remove the derivatives (recall the analogous
analysis for the spinless particle) which can be done at the cost
of introducing auxiliary Degrees Of Freedom (DOF). Simply stated
we will work in the Polyakov form since classically Nambu-Goto and
Polyakov forms are equivalent (on shell). Furthermore we will
exploit a trick, introduced by Lukierski et.al. \cite{luk1}, where
one introduces auxiliary variables, identified with time
derivatives of physical variables, to reduce or simplify the
overall time derivative structure of the Lagrangian. Indeed this
prescription does not lead to the simplified form in a unique way
so we will consider a form that is convenient for our purpose. (Of
course, for consistency of the scheme of \cite{luk1}, different
explicit forms, obtained due to this ambiguity, are physically
equivalent.)

Hence instead of (\ref{lag}) let us consider the classically
equivalent form,
\begin{equation}
L=p_\mu(\dot{x}^\mu - y^\mu)-\frac{y^2}{2e}-\frac{ly^2}{2e}
\frac{\sqrt{-\dot{k}^2}}{(k y)} - \frac{em^2}{2}+\lambda k^2.
\label{l2}
\end{equation}
Following \cite{luk1} we have replaced $\dot{x}^\mu$ by $y^\mu$
and this identification is enforced by the first term with $p_\mu
$ acting as a Lagrange multiplier. From (\ref{l2}) the original
from (\ref{lag}) can be recovered by integrating out $e$. The
advantage of extending the space of variables is that now the
Lagrangian is tractable for constraint analysis.

We now have on our hand a first order form of Lagrangian (\ref{l2})
where all the degrees of freedom are to be treated as independent
variables. The long list of conjugate momenta reads:
\begin{equation}
p_\mu^x=\frac{\partial L}{\partial \dot{x}^\mu}=p_\mu~~,
~~p_\mu^p=0~~,~~p_\mu^y=0~~,~~p^e=0~~,~~p^\lambda=0~~,
~~p_\mu^k=\frac{ly^2}{2e} \frac{\dot{k}_\mu}{(k
y)\sqrt{-\dot{k}^2}}. \label{mom}
\end{equation}
The canonical Hamiltonian $H$ follows,
\begin{equation}
H~=~p_\mu^x \dot{x}^\mu + p_\mu^p \dot{p}^\mu + p_\mu^y
\dot{y}^\mu +p_\mu^k \dot{k}^\mu + p^e \dot{e} + p^\lambda
\dot{\lambda} - L ~= ~(p y) + \frac{y^2}{2e} + \frac{em^2}{2} -
\lambda k^2.
\end{equation}

Notice that in (\ref{mom}) only $\dot k_\mu $ and no other
velocities appear. Hence majority of the velocities remain
undetermined meaning that there are constraints in the system.
This is not surprising since we have converted the $L$ of
(\ref{lag}) to a first order one in (\ref{l2}).

Before proceeding further we make a quick digression to discuss
the relevant points of Dirac's Hamiltonian analysis of constrained
systems \cite{dir}.\\
{\bf{Dirac's constraint analysis}}: ~~In Hamiltonian formulation,
any relation between dynamical variables, {\it{independent of
velocities}}, is considered as a constraint. Constraints can
appear directly in defining the conjugate momenta, (as in
(\ref{mom}) in the present case). New constraints can also be
generated from demanding time persistence of the first set of
constraints.

In the full set of constraints, the ones that commute with all
others(in the sense of Poisson brackets) are termed as First Class
Constraints (FCC). Rest of the non-commuting constraints are
termed as Second Class Constraints (SCC). The FCCs and SCCs have
to be treated in essentially different ways, especially if the system is being quantized.\\
Presence of constraints signifies a redundancy in the number of
DOF involved. FCCs induce presence of local gauge invariance. FCCs
can be treated in two ways. One can keep all the DOFs and impose
the FCCs by restricting the set of physical states to those
satisfying $(FCC)\mid state
>=0$. On the other hand, one can choose additional
constraints, (one each for one FCC), known as gauge fixing
conditions so that, these together with the FCCs turn in to an SCC
set.\\
For SCCs, a similar relation as above, $(SCC)\mid state
>=0$ can not be implemented consistently and one needs to modify the symplectic structure.
Poisson brackets have to replaced by Dirac bracket. Between two
generic variables $A$ and $B$ it is defined as,
\begin{equation}
\{A,B\}_{DB}=
\{A,B\}-\{A,(SCC)_i\}\{(SCC)_i,(SCC)_j\}^{-1}\{(SCC)_j,B\},\label{di}
\end{equation}
where $(SCC)_i$ is a set of SCC and $\{(SCC)_i,(SCC)_j\}$ is the
constraint matrix. Upon quantization, the Dirac brackets are
elevated to quantum commutators. For SCCs this matrix is
invertible. After exploiting the SCCs strongly as operator
relations and working with Dirac brackets one can work in a
reduced phase space, (as we will discuss in {\bf{section 3}}),
where SCCs are used to eliminate some DOF in favor of others.

The relevance of constraint analysis and Dirac brackets in the
context of NC geometry lies in the fact that the constraints
present in the model induce Dirac brackets that determine the NC
phase space algebra.

It is time now to consider our model. The first batch of
constraints, (known as Primary constraints \cite{dir}), appear
directly from (\ref{mom}):
\begin{equation}
\psi^{(1)}_{ \mu} \equiv p_\mu^x - p_\mu~~, ~~\psi^{(2)}_{ \mu}
\equiv p_\mu^p~~, ~~\psi^{(3)}_{ \mu} \equiv p_\mu^y~~,
~~\psi^{(4)} \equiv p^e~~,$$$$\psi^{(5)} \equiv
p^\lambda~~,~~\psi^{(9)} \equiv (p^k)^2+\frac{l^2 (y^2)^2}{4e^2(k
y)^2}. \label{c1}
\end{equation}
The last one, $\psi^{(9)}$ follows from ``squaring'' the last
relation in (\ref{mom}) involving $p_\mu ^k$. (This is similar to
deriving the mass-shell constraint from the Nambu-Goto form of
free spinless particle Lagrangian.) Invariance under time
translation yields the remaining Secondary constraints \cite{dir},
\begin{equation}
\dot \psi^{(3)}_{ \mu}=\{\psi^{(3)}_{ \mu},H \} \equiv
\psi^{(6)}_{ \mu} =  p_\mu + \frac{y_\mu}{e}~~,
~~\dot{\psi}^{(4)}+(p^\mu-\frac{y^\mu}{e})\psi^{(6)}_{ \mu}\equiv
\psi^{(7)}  =(p^x)^2-m^2~~,
$$$$ \dot \psi^{(5)}\equiv \psi^{(8)} = k^2~~,
~~ \dot{\psi}^{(9)}\equiv \psi^{(10)} = (k p^k). \label{c2}
\end{equation}
We use the convention for Poisson bracket and the metric as,
$$\{x_\mu,p^x_\nu\}=-g_{\mu \nu}~;~\{k_\mu,p^k_\nu\}=
-g_{\mu \nu} ~;~ g_{\mu \nu}=diag(1,-1,-1,-1).$$ Note that time
derivatives of $\psi^{(1)}_{ \mu}$ and $\psi^{(2)}_{\mu}$ are not
considered since this is a trivial set of SCC which is solved at
once and $p_\mu$ is replaced by $p_\mu^x $. This will not change
the algebra  the remaining DOF. Again $\psi^{(5)}$ is a trivial
FCC which we remove by fixing a gauge $\lambda=1$.

After trimming down the set of constraints to the non-trivial ones
we must make the essential classification of FCCs and SCCs among
them. For convenience we simplify and rename the constraints as $
\eta^{(1)},~\eta^{(2)},~\eta^{(3)}$ for
$\psi^{(7)},~\psi^{(8)},~\psi^{(10)} $ and $\phi^{(1)}_{
\mu},~\phi^{(2)}_{ \mu},~\phi^{(3)},~\phi^{(4)}$ for $
\psi^{(3)}_{ \mu},~\psi^{(6)}_{ \mu},~\psi^{(4)},~\psi^{(9)}$ so
that we have the following sets:
\begin{equation}
\eta^{(1)} \equiv (p^x)^2 -m^2~~,~~\eta^{(2)} \equiv k^2~~,
~~\eta^{(3)} \equiv (k p^k), \label{fcc}
\end{equation}
\begin{equation}
 \phi^{(1)}_{ \mu} \equiv p_\mu^y~~,~~\phi^{(2)}_{ \mu}
 \equiv p^x_\mu + \frac{y_\mu}{e}~~,~~\phi^{(3)} \equiv p^e~~,
 ~~\phi^{(4)} \equiv (p^k)^2+\frac{m^2 l^2 y^2}{4(k y)^2}.
\label{scc}
\end{equation}
One can directly check that the set $\eta $ in (\ref{fcc}) are FCC
that is they satisfy a closed algebra with all the constraints
$\eta, \phi $. From the $\phi $ set in (\ref{scc}) one can
construct two combinations,
\begin{equation}
\eta^{(4)} \equiv e \phi^{(3)} + (y \phi^{(1)})~;~~ \eta^{(5)}
\equiv \phi^{(4)} + \frac{e m^2 l^2 y^2}{2(k y)^3}(k \phi^{(2)}) -
\frac{e m^2 l^2}{2(k y)^2}(y \phi^{(2)}), \label{fcc1}
\end{equation}
that are also FCC. Below we provide the non-abelian FCC algebra of the set
$\eta $ in ((\ref{fcc}),(\ref{fcc1})):
\begin{equation}
\{\eta^{(2)} , \eta^{(3)}\} = -2 \eta^{(2)}~~, ~~\{\eta^{(2)} ,
\eta^{(5)}\} = -4 \eta^{(3)}~,
$$$$\{\eta^{(3)} , \eta^{(5)}\} = -2 \eta^{(5)}~~,
~~\{\eta^{(4)} , \phi^{(1)}_\mu\}=-\phi^{(1)}_\mu~.
\end{equation}
The remaining SCC set satisfies,
\begin{equation}
 \{\phi^{(1)}_{\mu},\phi^{(2)}_{\nu} \} = \frac{1}{e}g_{ \mu \nu}.
\label{scc1}
\end{equation}
We can solve this SCC system and the subsequent analysis will
remain unaffected. We find that the canonical Hamiltonian $ H=(p^x
y) + \frac{y^2}{2e}+\frac{em^2}{2}$ vanishes and hence the
dynamics will be governed by the FCCs only as is the case of
reparametrization invariant theories. Hence the extended
Hamiltonian, in  $e=1$ gauge, is,
\begin{equation}
H=\lambda_1 ((p^x)^2-m^2)+\lambda_2 k^2+ \lambda_3 (k
p^k)+\lambda_4 \left((p^k)^2 + \frac{m^4 l^2}{4 (k p^x)^2}\right),
\label{hfcc}
\end{equation}
where $\lambda_i $ are undetermined multipliers. It is
straightforward to check that the  constraint with $\lambda _4$,
in conjunction with the other constraints, reproduces the
Pauli-Lubanski condition in (\ref{pw}).

As we have mentioned in the discussion on Dirac procedure
\cite{dir} there are two ways of tackling a system with FCC. In
the gauge invariant formulation no gauge is fixed and one works
with all the DOF. We study this scheme in this section. In the
next section, {\bf{section 3}}, we will analyze the gauge fixed
version, which, incidentally is our main concern.

We now need to fix the multiplier $\lambda_i $. For this let us
follow Gitman and Tyutin \cite{gt}. The idea is to determine
$\lambda_i $ by comparing the expressions for velocities obtained
from (\ref{hfcc}),
\begin{equation}
\dot x_\mu =\{x_\mu ,H\}=-2\lambda _1p^x_\mu +\lambda
_4\frac{m^4l^2k_\mu}{2(kp^x)^3}, \label{01}
\end{equation}
and from the original action (\ref{lag}),
\begin{equation}
p^x_\mu =(\partial L)/(\partial \dot x^\mu ).
\end{equation}
Solving the above relations we find,
\begin{equation}
\lambda _1 =(\sqrt{\dot{x}^2})/\left(2m\sqrt{1+l\sqrt{
\frac{(-\dot{k}^2)}{(k \dot x)^2}}}\right)~~;~~ \lambda _4=-l(k\dot
x)\sqrt{(-\dot k^2)\left(1+l\sqrt{ \frac{(-\dot{k}^2)}{(k \dot
x)^2}}\right)}.
\end{equation}
Similarly for the other DOF $k_\mu $ we compare,
\begin{equation}
\dot k_\mu =\{k_\mu ,H\}=-\lambda _3k_\mu -2\lambda _4p^k_\mu,
\label{02}
\end{equation}
with
\begin{equation}
p^k_\mu =(\partial L)/(\partial \dot k^\mu )
\end{equation}
to obtain
\begin{equation}
\lambda _3 =0.
\end{equation}
Hence $\lambda_2$ remains arbitrary. In determining the
multipliers we have used the Dirac brackets obtained from the SCC
~~$\phi ^{(1)}_\mu ,~\phi ^{(2)}_\mu $~ which, however, do not
alter the brackets needed to compute ((\ref{01}),(\ref{02})). This
is because the set of SCC $\phi $ can modify only brackets of the
generic form $\{y_\mu ,~\}$ but the $H$ in (\ref{hfcc}) does not
contain $y_\mu $ {\footnote{See the comment below (\ref{scc}).}.

As a demonstration of the correctness of our constraint analysis
and the (Dirac) classification of constraints, one can go back to
((\ref{01}),(\ref{02})), substitute  $\lambda_i $ explicitly to
check that the equations of motion for $\dot x_\mu ,~\dot
k_\mu,~\dot{p}^x_\mu,~\dot{p}^k_\mu $ are identical with those of
\cite{wi}.

\section {\bf{New Non-Commutative Spacetime:}}
In this section our aim is to construct a minimal or
{\it{reduced}} NC phase space by exploiting the constraint
relations to eliminate some of the DOF. This is possible if one
introduces gauge fixing constraints for the FCCs and compute Dirac
brackets for the complete SCC system, consisting of FCCs
~~$\eta^{(1)}-\eta^{(5)}$~, gauge fixing constraints to be given
and the original SCCs $\phi ^{(1)}_\mu ,\phi ^{(2)}_\mu $. One can
perform a quick counting to check how many DOFs survive a full
gauge fixing. In the first order Lagrangian (\ref{l2}) we have the
phase space variables $ x_\mu ,k_\mu, y_\mu,e,
p_\mu,p^x_\mu,p^k_\mu,p^y_\mu,p^e$ meaning $26$ variables in
total. Using  $5$ FCC and $8$ SCC we can remove in all $18$
variables so that we are left with $8$ DOF in phase space that can
have dynamics. We might choose these to be $x_\mu,p^x_\mu $. This
means that, in principle, we can finally have a Hamiltonian
consisting of $x_\mu,p^x_\mu $ only and Dirac brackets for
$x_\mu,p^x_\mu $ the latter being the NC phase space for the RSPM
of \cite{st} that we have been advertising.

Indeed, the gauge fixing process is not unique and one fixes gauge
conditions according to the specific model in question as well as
the particular goal one has in mind. Our aim is to project this
RSPM as an extension to the spinless relativistic particle. For
the latter one has ~$L \sim m\sqrt{\dot x^2} $ leading to a single
FCC ~$\eta \sim (p^x)^2-m^2$. From here the gauge fixing for
reparametrization invariance (to fix the time variable) and
subsequent quantization can follow the analysis given in
\cite{gt}. This means that in the present case, out of the $5$
FCCs $\eta _1-\eta _5$ we will fix all but the first one, $\eta
^{(1)}=(p^x)^2-m^2$. This will make formally the Hamiltonian of
the RSPM same as the spinless one and all the complexity will be
present in the NC phase space algebra. Of course these NC Dirac
brackets will reduce to the canonical Poisson brackets for $l=0$.
Hence $l$ is to interpreted as the NC parameter.

Furthermore, this particular gauge system that we have introduced has another
important property: It allows a simple structure of Lorentz generator even in the
very complicated NC manifold and all the dynamical phase space variables
transform covariantly once the NC algebra is used.

In the present model a non-trivial technical problem arises in the
explicit gauge fixing process since our aim is to keep $\eta _1$
as a FCC so that one can fix time in the conventional way
\cite{gt}. This means that the gauge conditions have to be chosen
such that they commute with $\eta _1$.  The gauge fixing and the
resulting Dirac bracket computation is quite involved. In the
{\bf{Appendix}} we give an outline of the gauge fixing procedure
and a particular gauge fixing for which we now provide the Dirac
brackets. Note that we will provide the NC algebra in terms of the
variables $x_\mu,~p^x_\mu,~k_\mu,~p^k_\mu $ although we have fixed
all the FCC except $\eta _1$. This means that in principle we
could have given only the algebra consisting of $x_\mu,~p^x_\mu $
(without $k_\mu,~p^k_\mu $) but that will possibly be a very
complicated algebra.

Below we provide the NC phase algebra (or Dirac brackets) for RSPM
\cite{st} in a specific set of gauges (see {\bf{appendix}} for
details) is. First comes the $x_\mu ,p^x_\mu $ sector:
\begin{equation}
\{x_{\mu},x_{\nu}\}_{DB}=\frac{m^2
l^2(x_{\mu}k_{\nu}-k_{\mu}x_{\nu})}{4c^2(x
p^k)}~;~~\{p^x_{\mu},p^x_{\nu}\}_{DB}=0~,$$$$~~~\{x_{\mu},p^x_{\nu}\}_{DB}=-g_{\mu
\nu}-\frac{m^2 l^2(m^2 k_{\mu}k_{\nu}-c k_{\mu}p^x_{\nu})}{4c^3(x
p^k)} \label{ncx}
\end{equation}
Then we give the $k_\mu ,p^k_\mu $ Dirac brackets,
\begin{equation}
\{k_{\mu},k_{\nu}\}_{DB}=\frac{c (k_{\nu} p^k_{\mu}-k_{\mu}
p^k_{\nu})}{m^2(x p^k)}~,$$$$
\{p^k_{\mu},p^k_{\nu}\}_{DB}=-\frac{m^4 l^2(x_{\mu}
p^x_{\nu}-x_{\nu} p^x_{\mu})}{4 c^3(x p^k)}
 +\left(\frac{1}{c}-\frac{m^2 l^2}{4 c^2(x
p^k)}\right)(p^x_{\mu} p^k_{\nu}-p^k_{\mu} p^x_{\nu}) $$$$
+\frac{m^4 l^2}{4 c^4(x p^k)}(m^2(k_{\mu} x_{\nu} - x_{\mu}
k_{\nu})-(x p^x)(k_{\mu} p^x_{\nu} - k_{\nu} p^x_{\mu})),$$$$
\{k_{\mu},p^k_{\nu}\}_{DB}=-g_{\mu \nu}+\frac{m^2 p^k_{\mu}
x_{\nu}-(xp^x)p^k_{\mu} p^x_{\nu}}{m^2(x p^k)} +
\frac{1}{c}(k_{\mu} p^x_{\nu} -k_{\nu} p^x_{\mu})-\frac{m^2 l^2
k_{\mu} p^x_{\nu}}{4 c^2(x p^k)}-\frac{c p^k_{\mu}
p^k_{\nu}}{m^2(x p^k)}. \label{nck}
\end{equation}
Finally we have the mixed brackets,
\begin{equation}
\{x_{\mu},k_{\nu}\}_{DB}=-\frac{((x p^x) k_{\mu}+ c
x_{\mu})p^k_{\nu}}{m^2(x p^k)}-\left(-\frac{1}{c} + \frac{m^2
l^2}{4 c^2(x p^k)}\right) k_{\mu} k_{\nu}~, $$$$
\{x_{\mu},p^k_{\nu}\}_{DB}=-\frac{m^2 l^2(m^2 k_{\mu} x_{\nu}+c
x_{\mu} p^x_{\nu})}{4 c^3(x p^k)}-\frac{1}{c}(k_{\nu}
p^k_{\mu}+k_{\mu} p^k_{\nu})+\frac{m^2 l^2 k_{\mu} p^k_{\nu}}{4
c^2 (x p^k)}-\frac{m^4 l^2((x p^x) k_{\mu}k_{\nu}+c x_{\mu}
k_{\nu})}{4 c^4(x p^k)}, $$$$ \{k_{\mu},p^x_{\nu}\}_{DB}=\frac{m^2
p^k_{\mu}k_{\nu}-c p^k_{\mu} p^x_{\nu}}{m^2(x p^k)} $$$$
\{p^x_{\mu},p^k_{\nu}\}_{DB}=-\frac{m^2 l^2(m^2 k_{\mu}
p^x_{\nu}-c p^x_{\mu} p^x_{\nu})}{4 c^3(x p^k)}+\frac{m^4 l^2
k_{\nu} p^x_{\mu}}{4 c^3(x p^k)}-\frac{m^6 l^2 k_{\mu} k_{\nu}}{4
c^4 (x p^k)} $$$$ . \label{xk}
\end{equation}
In the above NC algebra $c\ne 0$ is a numerical parameter that
appears from gauge fixing (see {\bf{Appendix}}).

One point regarding the above NC algebra
(\ref{ncx},\ref{nck},\ref{xk}) should be mentioned. For $l=0$ the
$x_\mu ,p^x_\mu $ sector reduces to the canonical one but rest of
the brackets, although simplifies considerably, still remain
non-trivial.

The above NC algebra can be put to a non-trivial use: generating
the Lorentz algebra from the covariant angular momentum operator
$M_{\mu\nu}$. It is important to note that using
((\ref{ncx}),(\ref{nck}),(\ref{xk})),  algebra among $x_{\mu}
p^x_{\nu} - x_{\nu} p^x_{\mu} $, (which is the angular momentum
for the spinless particle), does not close. But the correct
Lorentz algebra,
\begin{equation}
\{M_{\mu \nu},M_{\alpha \beta}\}_{DB}= g_{\mu \beta} M_{\nu
\alpha} + g_{\nu \alpha} M_{\mu \beta} + g_{\mu \alpha} M_{\beta
\nu} + g_{\nu \beta} M_{\alpha \mu},
\end{equation}
is recovered only when the angular momentum has a spin
contribution. The correct generator is{\footnote{The generator
$M_{\mu\nu}$ is gauge invariant which can be seen easily because
it commutes with all the FCC listed in the expression for the
Hamiltonian in (\ref{hfcc}), where one is free to use canonical
Poisson brackets. It is a conserved quantity as well.}},
\begin{equation}
M_{\mu \nu}=x_{\mu} p^x_{\nu} - x_{\nu} p^x_{\mu} + k_{\mu}
p^k_{\nu} - k_{\nu} p^k_{\mu}.
\end{equation}
The demonstration of the above is straightforward but quite
tedious and involves a number of non-trivial cancellations between
terms coming from different sectors. The structure of $M_{\mu
\nu}$ is quite elegant when contrasted to the more involved form
given in \cite{wi}.

As we have already advertised it is straightforward to explicitly
check, using the NC Dirac brackets (\ref{ncx},\ref{nck},\ref{xk}),
that this $M_{\mu \nu}$ reproduces the usual Lorentz
transformation for all the physical degrees of freedom:
\begin{equation}
\{M_{\mu \nu}, x_{\rho} \}_{DB}=g_{\nu \rho}x_\mu-g_{\mu
\rho}x_\nu~~,~~\{M_{\mu \nu}, p^x_\rho\}_{DB}=g_{\nu
\rho}p^x_\mu-g_{\mu \rho}p^x_\nu~,$$$$ \{M_{\mu \nu},
k_\rho\}_{DB}=g_{\nu \rho}k_\mu-g_{\mu \rho}k_\nu~~,~~\{M_{\mu
\nu}, p^k_\rho\}_{DB}=g_{\nu \rho}p^k_\mu-g_{\mu \rho}p^k_\nu.
\label{la}
\end{equation}

\section {\bf{Spinning Particle Dynamics:}}
We have entered the final stage of our analysis of the RSPM of
\cite{st} where we derive the Hamiltonian dynamics by fixing the
time in the canonical gauge (for details see \cite{gt}). The
essential point is that one rewrites the remaining FCC $\eta _1
=(p^x)^2-m^2$ as,
\begin{equation}
 \eta _1= \mid p_0 \mid -\sqrt{p^x_ip^x_i+m^2},
\label{h}
\end{equation}
and fix the gauge $x'_0=0$. This amounts to a gauge fixing of the
form $x'_0\sim x_0 -\tau =0$ with $(x_i)'=x_i,~(p_\mu )'=p_\mu $.
This requires $x_0$ to be a $c$-number parameter (time) and this
formal manipulation is convenient because one can then directly
apply the Dirac analysis with the time-independent constraint
$\sim x'_0  =0$ (we follow \cite{gt}). In the spinless case, with
this gauge choice one recovers the dynamics of physical variables
$\dot x_i=p_i/\sqrt{p^x_ip^x_i +m^2}$ from the physical reduced
space Hamiltonian $H=\sqrt{p^x_ip^x_i +m^2}$ using Dirac brackets following
from the SCCs $x'_0~,~\mid p_0 \mid -\sqrt{p^x_ip^x_i+m^2}$.

In the present case also we need to carry out the above procedure
keeping in mind the last iteration of Dirac brackets
((\ref{ncx}),(\ref{nck}),(\ref{xk})). Specifically we consider the
FCC and gauge fixing,
\begin{equation}
 \eta _1= \mid p'_0 \mid -\sqrt{(p^x)'_i(p^x)'_i+m^2}~,~~x'_0=0,
\label{h1}
\end{equation}
(where it is simply a renaming for $(x_i)'=x_i,~(p_\mu )'=p_\mu
$), and compute the last stage Dirac brackets using the
penultimate stage Dirac brackets (\ref{ncx},\ref{nck},\ref{xk}).
The final NC algebra for the physical $x_i,p^x_i$ sector is,
\begin{equation}
 \{x_i,x_j\}_{DB}=\frac{l^2m^2}{4c^2(xp^k)}(x_i(k_j+\frac{k_0p_j}
 {\sqrt{\vec p^2+m^2}})-x_j(k_i+\frac{k_0p_i}{\sqrt{\vec p^2+m^2}})),$$$$
\{x_i,p_j\}_{DB}=-g_{ij}-\frac{l^2m^2}{4c^2(xp^k)}(k_i+\frac{k_0p_i}{\sqrt{\vec
p^2+m^2}})(m^2k_j-cp_j)~;~\{p_i,p_j\}=0.
\end{equation}
Note that the final Dirac brackets  in the
$k_\mu,~p^k_\mu,~p^x_\mu $ sector remains {\it{unchanged}} and is
same as (\ref{nck}). This occurs due to $\{\eta _1,k_\mu
\}=0;~\{\eta _1,p^k_\mu \}=0;~\{\eta _1,p^x_\mu \}=0$ using
(\ref{xk}) and this calculation is quite amusing. However there
will again be changes in the mixed $x_\mu ,k_\mu $ and $x_\mu
,p^k_\mu $ sectors which we have not shown here.

Finally we recover the cherished forms of the equations of motion
for the free RSPM \cite{st,wi}:
\begin{equation}
\dot x_i=-\frac{p_i}{\sqrt{\vec
p^2-m^2}}-\frac{l^2m^2(m^2k_0-c\sqrt{\vec p^2-m^2})}
{4c^3(xp^k)}(k_i+\frac{k_0}{\sqrt{\vec p^2-m^2}}p_i), \label{ex}
\end{equation}

\begin{equation}
\dot p_i =0. \label{ep}
\end{equation}
For consistency one can check that the same result is obtained
either from $\dot x_i=\{x_i,p_0\}_{DB}$ or from $\dot
x_i=\{x_i,\sqrt{\vec p^2-m^2}\}_{DB}$. The dynamics of $k_\mu $
can be obtained straight from (\ref{xk}),
\begin{equation}
\dot k_\mu=\frac{(m^2k_0-c\sqrt{\vec p^2-m^2})}{m^2(xp^k)}p_\mu
^k. \label{ek}
\end{equation}

Another interesting finding is the presence of a conserved
pseudovector,
\begin{equation}
G_i=\epsilon_{ijl}(m^2k_j-cp^x_j)p^k_l~;~~\dot G_i=0. \label{g}
\end{equation}
The equations of motion and the conserved vector $G_i$ are the
Hamiltonian analogues of the the Lagrangian equations of motion
obtained in \cite{wi}.
\section {Concluding Remarks and Future Prospects}}
The relativistic spinning particle model, proposed in
\cite{kuz,st} has several interesting features that need further
careful investigation. The main interest lies in the fact the
model is a classical one but nevertheless has quantum particle
like property as regards to its spin. The spin appears as a
Casimir and hence possess a universal character. From a purely
mathematical point of view this extension of the standard spinless
particle model is quite novel and has origin in group theoretic
ideas. Many interesting behaviors of the particle have been
revealed in the Lagrangian analysis conducted by Kassandrov et.al.
\cite{wi}.

In the present work we have carried a thorough Hamiltonian
analysis of the model. The principal result of our work is the
emergence of a new form of non-commutativity in an (extended)
spacetime. Using this non-commutative phase space approach we have
corroborated our results for the particle dynamics with those of
\cite{wi}. Furthermore, even if it is treated only as a problem of
interest in constrained dynamical system, the Hamiltonian Dirac
constraint analysis possesses several subtle and intricate
features.

There are a number of avenues open for further study on this model
in our proposed Hamiltonian framework. The non-commutative algebra
needs to be  analyzed carefully since there might be  combinations
of the fundamental degrees of freedom ($x_\mu,p^x_\mu, k_\mu,
p^k_\mu $) that can lead to a simpler non-commutative symplectic
structure which will help to understand the dynamics better. The
other (hopefully straightforward) task is to include external
interactions in the model along the lines of \cite{wi}.

Finally, Hamiltonian formulation is the starting point from which
one can attempt a proper quantization of the model. In the present
case, the way we have formulated the problem, the Hamiltonian has
reduced to that of a simple free spinless particle and all the
spin related complexity resides in the non-commutative phase space
algebra. A similar thing appears in the previous spinning particle
models \cite{hr,sg1} as well. Hence a non-trivial job is to find,
at least to lowest order in non-commutativity, a proper
representation of the classical variables as quantum operators.
Once again, the first step towards this realization lies in
deriving a (Darboux like) map from the non-commutative degrees of
freedom to canonical degrees of freedom, in the classical setup.
This will probably lead to a more complicated Hamiltonian, with
spin effects included, but a simpler canonical phase space
algebra, so that a perturbative Schrodinger analysis may be
performed in a non-relativistic approximation. \vskip .3cm
\noindent
 {\bf{Appendix}}:\\
We outline computation of the Dirac brackets
(\ref{ncx},\ref{nck},\ref{xk}) resulting in the NC phase space.
For a specific system with a large number of SCCs the Dirac
brackets can be evaluated iteratively that is one can start by,
say, any two SCCs, compute their Dirac bracket, then use this as
the starting bracket to compute Dirac brackets for rest of the
SCCs. Since in the present case we have three FCCs so to calculate
Dirac brackets for a system of six SCCs is difficult. Hence we
start with $\eta ^{(5)}$ (with $\phi^{(2)}_\mu $ removed) and fix
a gauge condition
\begin{equation}
\eta ^{(5)}_g \equiv \frac{1}{2}\left((x k) - \frac{(x p^x)(k
p^x)}{m^2}\right)=0. \label{g1}
\end{equation}
The constraint matrix for the SCC pair $\eta ^{(5)},~\eta
^{(5)}_g$ and its inverse are given by
\begin{equation}
\{\eta ^{(5)},~\eta ^{(5)}_g \}=
 \left (
\begin{array}{cc}
 0 &  a\\
-a &  0
\end{array}
\right ) \label{mat}~~~~,~~~~ \{\eta ^{(5)},~\eta ^{(5)}_g
\}^{-1}=
 \left (
\begin{array}{cc}
 0 &  b\\
-b &  0
\end{array}
\right )
\end{equation}
where $$a=(x p^k)-\frac{(p^k p^x)(x p^x)}{m^2}~;~~
b=\frac{m^2}{(p^k p^x)(x p^x)-m^2(x p^k)}.$$ This gives rise to
Dirac brackets which are not shown. Then we fix gauges
\begin{equation}
\eta ^{(3)}_g  \equiv (k p^x)-c =0~;~~\eta ^{(2)}_g \equiv (p^k
p^x) =0, \label{g2}
\end{equation}
for the remaining two FCCs and again calculate Dirac brackets
using the previously obtained Dirac brackets as the starting phase
space algebra. In (\ref{g2}) ~$c$~ is a non-zero numerical
parameter. It should be mentioned that the specific gauge choices
we have made are dictated by our target of keeping the FCC nature
of $\eta ^{(1)}$ intact. Indeed, one is free to make other
(allowed) gauge choices. The SCC constraint matrix consisting of
brackets among $ \eta _3,\eta ^{(3)}_g,\eta _2,\eta ^{(2)}_g $ is
now  $4$-dimensional. Below the constraint matrix $[matrix]_{ij}$
is written  with the constraints arranged in order with
$[matrix]_{11}=\{\eta _3, \eta _3 \}$.
\begin{equation}
[matrix]_{ij}= \left (
\begin{array}{cccc}
 0 & a_1 & 0 & 0\\
 -a_1 & 0 & 0 & -a_2\\
 0 & 0 & 0 & -2 a_1\\
 0 & a_2 & 2 a_1 & 0
\end{array}
\right ) \label{mat1}~~~~,~~~~ [matrix]^{-1}_{ij}=
 \left (
\begin{array}{cccc}
 0 & -b_1 & b_2 & 0\\
 b_1 & 0 & 0 & 0\\
 -b_2 & 0 & 0 & -b_1 /2\\
 0 & 0 & b_1 /2 & 0
\end{array}
\right )
\end{equation}
where $$a_1=(k p^x)~,~a_2=m^2 + \frac{m^6 l^2}{4(k p^x)((p^k
p^x)(x p^x)-m^2(x p^k))}$$  $$b_1=1/(k p^x)~,~b_2=\left(m^2 +
\frac{m^6 l^2}{4(k p^x)((p^k p^x)(x p^x)-m^2(x p^k))}\right) / 2(k
p^x)^2.$$ From here the Dirac brackets
(\ref{ncx},\ref{nck},\ref{xk}) are computed. \vskip .4cm \noindent
{\bf{Acknowledgements:}} We are very grateful to Professor A.
Staruszkiewicz for promptly sending us his paper upon our request.
We thank Professor S.M. Kuzenko for informing us of their earlier
work \cite{kuz}. We also thank Professor V. Kassandrov and
Professor P. Horvathy for correspondence.

\end{document}